%% file: main.tex
\documentclass[journal=nalefd,manuscript=letter,layout = standard]{achemso}
\usepackage{chemformula} 
\usepackage[T1]{fontenc} 
\usepackage{multirow}
\usepackage{hyperref}
\usepackage[uncertainty-mode = separate, separate-uncertainty-units = single, range-phrase=--, range-units = single, detect-all]{siunitx}
\usepackage[utf8]{inputenc}
\usepackage{amssymb}

\author{Christine Falter}
\email{c.falter@fz-juelich.de}
\affiliation[RWTH]{JARA-Fundamentals of Future Information Technology, Jülich-Aachen Research Alliance, Forschungszentrum Jülich and RWTH Aachen University, Germany}
\alsoaffiliation[PGI]{Peter-Grünberg-Institut 9, Forschungszentrum Jülich GmbH, 52428 Jülich}

\author{Yurii Kutovyi}
\affiliation[PGI]{Peter-Grünberg-Institut 10, Forschungszentrum Jülich GmbH, 52428 Jülich}
\alsoaffiliation[RWTH]{JARA-Fundamentals of Future Information Technology, Jülich-Aachen Research Alliance, Forschungszentrum Jülich and RWTH Aachen University, Germany}

\author{Nils von den Driesch}
\affiliation[PGI]{Peter-Grünberg-Institut 10, Forschungszentrum Jülich GmbH, 52428 Jülich}
\alsoaffiliation[RWTH]{JARA-Fundamentals of Future Information Technology, Jülich-Aachen Research Alliance, Forschungszentrum Jülich and RWTH Aachen University, Germany}

\author{Denny Dütz}
\affiliation[RWTH]{JARA-FIT Institute for Quantum Information, Forschungszentrum J\"ulich GmbH and RWTH Aachen University, 52074 Aachen, Germany}

\author{Nataliya Demarina}
\affiliation[RWTH]{JARA-Fundamentals of Future Information Technology, Jülich-Aachen Research Alliance, Forschungszentrum Jülich and RWTH Aachen University, Germany}
\alsoaffiliation[PGI]{Peter-Grünberg-Institut 2, Forschungszentrum Jülich GmbH, 52428 Jülich}

\author{Prateek Kaul}
\affiliation[RWTH]{JARA-Fundamentals of Future Information Technology, Jülich-Aachen Research Alliance, Forschungszentrum Jülich and RWTH Aachen University, Germany}
\alsoaffiliation[PGI]{Peter-Grünberg-Institut 9, Forschungszentrum Jülich GmbH, 52428 Jülich}

\author{Anjana Rajan}
\affiliation[RWTH]{JARA-Fundamentals of Future Information Technology, Jülich-Aachen Research Alliance, Forschungszentrum Jülich and RWTH Aachen University, Germany}
\alsoaffiliation[PGI]{Peter-Grünberg-Institut 9, Forschungszentrum Jülich GmbH, 52428 Jülich}

\author{Jiayuan Zhang}
\affiliation[RWTH]{JARA-Fundamentals of Future Information Technology, Jülich-Aachen Research Alliance, Forschungszentrum Jülich and RWTH Aachen University, Germany}
\alsoaffiliation[PGI]{Peter-Grünberg-Institut 9, Forschungszentrum Jülich GmbH, 52428 Jülich}

\author{Thomas J. Smart}
\affiliation[RWTH]{JARA-Fundamentals of Future Information Technology, Jülich-Aachen Research Alliance, Forschungszentrum Jülich and RWTH Aachen University, Germany}
\alsoaffiliation[PGI]{Peter-Grünberg-Institut 9, Forschungszentrum Jülich GmbH, 52428 Jülich}

\author{Leqi Zhou}
\affiliation[RWTH]{JARA-Fundamentals of Future Information Technology, Jülich-Aachen Research Alliance, Forschungszentrum Jülich and RWTH Aachen University, Germany}
\alsoaffiliation[PGI]{Peter-Grünberg-Institut 9, Forschungszentrum Jülich GmbH, 52428 Jülich}

\author{Lidia Kibkalo}
\affiliation[ERC]{Ernst Ruska-Centre for Microscopy and Spectroscopy with Electrons, Forschungszentrum Jülich GmbH, 52428 Jülich}

\author{Andr\'as Kov\'acs}
\affiliation[ERC]{Ernst Ruska-Centre for Microscopy and Spectroscopy with Electrons, Forschungszentrum Jülich GmbH, 52428 Jülich}

\author{Qing-Tai Zhao}
\affiliation[RWTH]{JARA-Fundamentals of Future Information Technology, Jülich-Aachen Research Alliance, Forschungszentrum Jülich and RWTH Aachen University, Germany}
\alsoaffiliation[PGI]{Peter-Grünberg-Institut 9, Forschungszentrum Jülich GmbH, 52428 Jülich}

\author{Lars R. Schreiber}
\affiliation[RWTH]{JARA-FIT Institute for Quantum Information, Forschungszentrum J\"ulich GmbH and RWTH Aachen University, 52074 Aachen, Germany}
\alsoaffiliation[ARQUE]{ARQUE Systems GmbH, 52074 Aachen, Germany}

\author{Alexander Pawlis}
\email{a.pawlis@fz-juelich.de}
\affiliation[PGI]{Peter-Grünberg-Institut 9, Forschungszentrum Jülich GmbH, 52428 Jülich}
\alsoaffiliation[PGI]{Peter-Grünberg-Institut 10, Forschungszentrum Jülich GmbH, 52428 Jülich}
\alsoaffiliation[RWTH]{JARA-Fundamentals of Future Information Technology, Jülich-Aachen Research Alliance, Forschungszentrum Jülich and RWTH Aachen University, Germany}

\title[FET]
  {Realization of \ch{ZnSe}-based Field-Effect Transistors operating at Cryogenic Temperatures as a Platform for Future Spin-Qubit Applications}

\begin{document}

\newpage
\begin{abstract}
   The wide-bandgap compound semiconductor \ch{ZnSe} is a promising host material for the realization of electron spin-qubits. Its non-degenerate conduction band and the potential for isotopic nuclear spin purification promise long spin coherence times. Key requirements for such devices include reliable electrostatic control of electrons in \ch{ZnSe} and low-resistance ohmic contacts that remain functional at cryogenic temperatures. In this work, we utilize a novel Shadow Wall technique for molecular-beam epitaxy combined with in-situ deposition of \ch{Al} ohmic contacts to realize normally-off \ch{ZnSe}-based field-effect transistors. The devices exhibit linear output characteristics and effective gate control of the channel from room temperature down to \qty{5}{\K}, confirming low-resistance ohmic contacts to the undoped \ch{ZnSe} channel. The drain current can be modulated by several orders of magnitude through electrostatic gating, with threshold voltages of approximately \qty{3}{V} and field-effect mobilities exceeding $\qty{100}{cm^2 / V\cdot s}$ over the investigated temperature range. Self-consistent Schrödinger-Poisson and drift-diffusion simulations reproduce the measured transfer characteristics and provide insight into the role of interface electrostatics in determining the channel formation and threshold voltage. These results demonstrate the potential of gated \ch{ZnSe} heterostructures for future spin-qubit applications.
\end{abstract}

\newpage
\section{Introduction}
\input{chapters/introduction}

\section{Results and Discussion}
\subsection{Realization of FETs using the SWE technique}
\input{chapters/structural}
\subsection{Electrical Characterization}
\input{chapters/electrical}
\section{Conclusion}

\input{chapters/conclusion}
\newpage
\section{Methods}
\input{chapters/methods}

\begin{acknowledgement}
We gratefully acknowledge the technical support and service by B. Bennemann, C. Krause and from the staff of HNF at Forschungszentrum Jülich for their assistance with device fabrication. This work is supported by the German Research Foundation (DFG) within the project no. 337456818 and by Germany's Excellence Strategy - Cluster of Excellence Matter and Light for Quantum Computing (ML4Q) EXC 2004/2 - 390534769. The authors gratefully acknowledge the use of resources in the ''ER-C 2.0`` National Research Infrastructure for High-Resolution Electron Microscopy in Forschungszentrum J\"ulich GmbH.

\end{acknowledgement}

\section*{Conflict of Interest}
The authors declare no conflicts of interest.

\section*{Data Availability Statement}
The data is available from the authors upon reasonable request.

\bibliography{bibliography}

\end{document}

%% file: chapters/introduction.tex
Electron and hole spin-qubits hosted in gate-defined quantum dots in planar \ch{Si}/\ch{SiGe} and Ge/SiGe heterostructures are among the leading platforms for large-scale quantum computing \cite{Burkard2023, Scappucci2020}. These qubits can be controlled entirely electrically at high clock rates~\cite{defuentes2025, HRL2026, Dijkema2026}, while their manipulation, initialization, and readout fidelities have already surpassed the thresholds required for quantum error correction~\cite{Noiri2022, Mills2022, Xue2022, Wu2025, HRL2026}. Combined with conveyor-belt qubit shuttling~\cite{desmet2024, Xue2024, Ademi2025, Beer2026, Matsumoto2026}, modular and scalable architectures for quantum chips comprising millions of qubits~\cite{Boter2022, kunne2024spinbus}, compatible with industrial foundries~\cite{George2025, Muster2025Shuttling} have been proposed. However, material-related issues, including the conduction band valley splitting due to alloy disorder in \ch{SiGe}~\cite{klos_atomistic_2024, Volmer2026} and g-tensor disorder for hole spins caused by local strain variations ~\cite{Scappucci2020, Seidler2025} as well as sparse crystalline defects related to the relaxed buffer layers pose significant challenges for the realization of electrostatic spin-qubits. \\
\\
Concurrently, electron spin-qubits in the II/VI semiconductor \ch{ZnSe} would combine several key advantages of established semiconductor qubit platforms. In particular, \ch{ZnSe} possesses nuclear-spin-free isotopes~\cite{Kirstein2021,Pawlis2019}, which are crucial for achieving long electron spin coherence times\cite{Struck2020, HRL2026}. As a direct bandgap semiconductor~\cite{Continenza1988}, \ch{ZnSe} enables efficient optical spin-photon interfaces, while its non-degenerate conduction band eliminates complications associated with valley splitting. Furthermore, highly crystalline and defect-free \ch{ZnSe} heterostructures can be grown fully strained on \ch{GaAs} wafers via Molecular Beam Epitaxy (MBE)~\cite{Gunshor1988}, removing the presence of dislocations that emerge from relaxed virtual substrates.~\cite{Scappucci2020}. Finally, electrons in \ch{ZnSe} exhibit moderate spin-orbit interaction without g-tensor issues known for holes, enabling reliable electric-dipole spin resonance (EDSR) control without requiring micromagnets~\cite{Nowack2007}.\\
\\

The realization of gate-defined quantum dots in \ch{ZnSe} heterostructures requires precise electrostatic control of the carrier concentration and high-quality ohmic contacts that remain operational at cryogenic temperatures. Before more complex quantum-dot architectures can be implemented, these key technological requirements should first be demonstrated in a simple gate-controlled device. In this regard, field-effect transistors (FETs) provide an ideal test platform to evaluate the quality of the semiconductor heterostructure, the gate dielectric, and the electrical contacts, while simultaneously assessing the efficiency of electrostatic carrier modulation. Early \ch{ZnSe}-based transistors were based on MBE grown \ch{Cl}-doped \ch{ZnSe} deposited on Cr-doped GaAs to operate in depletion mode~\cite{Dreifus1990}. However, the realization of reliable, small area ohmic contacts was limiting in these devices. The formation of local ohmic contacts with low contact resistivity to \ch{ZnSe} layers, particularly at cryogenic temperatures, has proven to be a significant challenge in the \ch{ZnSe}-based system. Traditional methods based on implantation or annealing of the contact metal at high temperature are counterproductive, since the diffusion of the metal atoms can lead to the formation of p-type group II vacancies that degrade contact performance particularly at cryogenic temperatures~\cite{Kuttler1996, Jansen2020, Wu2022}. However, recently, contact resistivities as low as $\qty{e-5}{\Omega\cdot cm^2}$ have been achieved by \textit{in-situ} deposition of \ch{Al} on highly-doped ($\qty{e19}{cm^{-3}}$) n-type \ch{ZnSe:Cl} without the need of any post-growth annealing procedure~\cite{Jansen2020}. Furthermore, reliable operation and comparable low contact resistivities of these contacts at \qty{4}{K} was observed.\\
\\
In this work, we combine the \textit{in-situ} deposition of \ch{Al} ohmic contacts to n-type \ch{ZnSe:Cl} with a Shadow Wall Epitaxy (SWE) technique for MBE, that allows for the selective area growth of compound semiconductors and additionally, reduces the number of required post-fabrication steps~\cite{vondenDriesch2024}. Particularly in the \ch{ZnSe} system, abrupt epitaxially grown regions are achieved using SWE due to the low adatom diffusion length and sticking coefficients of \ch{Zn} and \ch{Se} species without the presence of their respective elemental partner~\cite{Riley1996,Schallenberg2003,vondenDriesch2024}. Here we demonstrate, that the SWE technique can be extended to allow for selective doping of the \ch{ZnSe} layer~\cite{PIN}, in conjunction to the \textit{in-situ} deposition and patterning of ohmic \ch{Al} contacts on top of these highly-doped regions to realize accumulation-mode \ch{ZnSe}-based FETs. \\
\\

Our results indicate, that the SWE-based fabrication approach is uniquely suited to realize high-quality \ch{ZnSe}-based heterostructures with no contact degradation at cryogenic temperatures. The output and transfer characteristics of the transistors are systematically investigated over a broad temperature range from \qty{295}{K} down to \qty{5}{K} to evaluate the contact performance, gate control, and carrier transport in the \ch{ZnSe} channel. The transistors exhibit stable normally-off operation from room temperature down to \qty{5}{\K}, demonstrating reliable electrostatic control of electron accumulation in the undoped \ch{ZnSe} channel together with low-resistance ohmic contacts.  Self-consistent Schrödinger-Poisson and drift-diffusion simulations reveal the electrostatics governing the formation of the conductive channel. These results establish the SWE process as a viable technology for the fabrication of gated \ch{ZnSe} heterostructures and provides an essential technological foundation for future gate-defined ZnSe spin-qubit devices.

%% file: chapters/structural.tex
For this work, \ch{ZnSe}-based FETs are developed using the SWE technique, which relies on the shadowing of directional material fluxes in MBE by vertical shadow walls~\cite{vondenDriesch2024}. To this end, we fabricated \ch{Si}-based shadow walls on top of \ch{ZnMgSe}-buffers that were prepared on epi-ready \ch{GaAs} (001) substrates and subsequently \textit{in-situ} passivated with \ch{AlO_x}. The height of these walls was approximately \qty{1.95}{\um}, consisting of \qty{150}{nm} \ch{SiO2} and \qty{1.8}{\um} \ch{Si}. The shadow walls were patterned utilizing a \ch{BaF2} hard mask and a combination of dry and wet etching techniques, described in further detail in the Methods section.
\\
\begin{figure}
    \centering
    \includegraphics[]{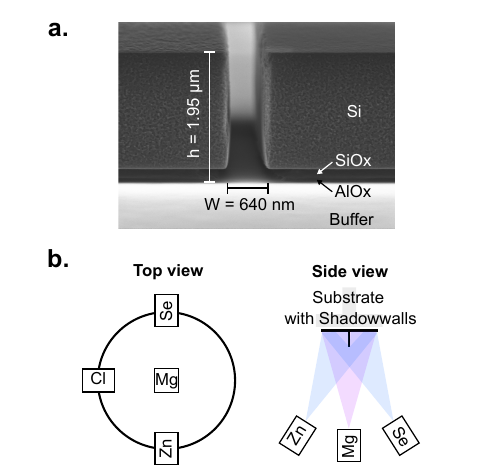}
    \caption{Realization of \ch{ZnSe}-based FETs using SWE. (a) SEM side-view of the completed shadow wall structure. (b) Schematic of the effusion cell arrangement within the II/VI MBE as viewed from the top of the chamber (left) and from the side (right). In the side-view image, the shadowing of the directional material fluxes on the substrate is illustrated.}
    \label{fig:structure}
\end{figure}
\\
Figure~\ref{fig:structure}~a shows a secondary electron (SE) scanning electron microscopy (SEM) micrograph of typical shadow walls just before re-introduction to the MBE system. Well-defined vertical walls are achieved by RIE-etching of the \ch{Si} with low anisotropy. During RIE-etching the \ch{ZnMgSe} buffer was protected by the remaining \ch{SiO2} as well as the underlying \ch{AlO_x} passivation layer. Both of these layers were removed in a second etching step using diluted hydrofluoric acid (HF). This wet etching step creates the lateral undercut visible in Fig.~\ref{fig:structure}~a. Importantly, the HF also selectively removes the \ch{BaF2} hardmask, exposing the sharp, well-defined \ch{Si} edges essential for SWE. To minimize the re-oxidation of the \ch{ZnMgSe} buffer surface, the HF etching was performed within 10\,minutes of transferring the sample into the load lock of the MBE system. For the FET heterostructure, a stack composed of \qty{20}{nm} undoped \ch{ZnMgSe} and \qty{50}{nm} highly-doped \ch{ZnSe:Cl} was deposited on top of the pre-grown \ch{ZnMgSe} buffer (see Methods for detailed growth process).\\
\\
The realization of a FET device using the SWE technique relies on the angle between the atom beams for \ch{Zn},\ch{Se} and \ch{Cl} as well as the orientation of the walls during epitaxial growth. An illustration of the cell configuration is provided in Fig.~\ref{fig:structure}~b both from the top (left image) and side (right image) of the MBE chamber. The \ch{Mg} cell is located at the bottom of the chamber such that the material flux is parallel to the substrate normal and therefore, no shadowing of the \ch{Mg}-flux occurs, regardless of sample orientation. By contrast, the \ch{Zn}, \ch{Se} and \ch{Cl} cells are aligned at an angle of about \qty{30}{°} to the substrate normal.  The FET channel is defined by a gap of width $W$ between two shadow walls (see Fig.~\ref{fig:structure}~a). During growth, the \ch{Zn} and \ch{Se} fluxes are aligned parallel to this gap while the \ch{Cl} flux is oriented perpendicularly. As a consequence, the \ch{Cl} beam is interrupted by the walls during the \ch{ZnSe:Cl} growth, such that the \ch{ZnSe} in between the two walls remains undoped, forming the FET channel. At the same time, the grown \ch{ZnSe} is highly doped in the source and drain contact regions at either end of the channel. In order to prevent the presence of \ch{Cl} in the channel region, $W$ must be smaller than the shadow length, i.e. $W<L_\text{Shadow}^{(II/VI)}\approx h\cdot\tan(30^\circ)=\qty{1.1}{\um}$, determined by the height $h=\qty{1.95}{\um}$ of the walls and the angle of the effusion cells with regards to the substrate normal.
This constraint limits the maximum achievable channel width and, consequently, the drain-source current. To mitigate this, we also implemented FETs with multiple parallel channels with the same width $W$ and length $L$. A device with $N$ identical parallel channels can be modeled as a single transistor with effective channel width $W_\text{eff.}=N\cdot W$.\\
\\
To form \textit{in-situ} low-ohmic contacts to the \ch{ZnSe} channel, the sample was transferred under UHV conditions to a separate MBE chamber, where \qty{60}{nm} of \ch{Al} was deposited on top of the highly doped \ch{ZnSe} region~\cite{Jansen2020}. The angle of the \ch{Al}-cell with regards to the samples surface normal is about \qty{32}{°}, leading to a shadow length of $L_\text{Shadow}^{(\ch{Al})}\approx\qty{1.2}{\um}$, similar to the II/VI shadow length. Therefore by aligning the \ch{Al} perpendicularly to the channel, we ensure that there is no \ch{Al} deposition in the channel region. Using this growth scheme, two distinct \ch{Al} contact pads at the ends of the channel were formed without the need for post-growth processing. Subsequently, the entire structure was covered with a conformal dielectric stack consisting of \qty{20}{nm} \ch{Al2O3} and \qty{5}{nm} \ch{HfO2} grown via atomic layer deposition (ALD). Finally, the gate contact comprising a  \qty{10}{nm} \ch{Ti} adhesion layer and a \qty{75}{nm}  \ch{Au} cap was defined using post-growth optical lithography and a lift-off process.\\
\\
\begin{figure*}
    \centering
    \includegraphics[]{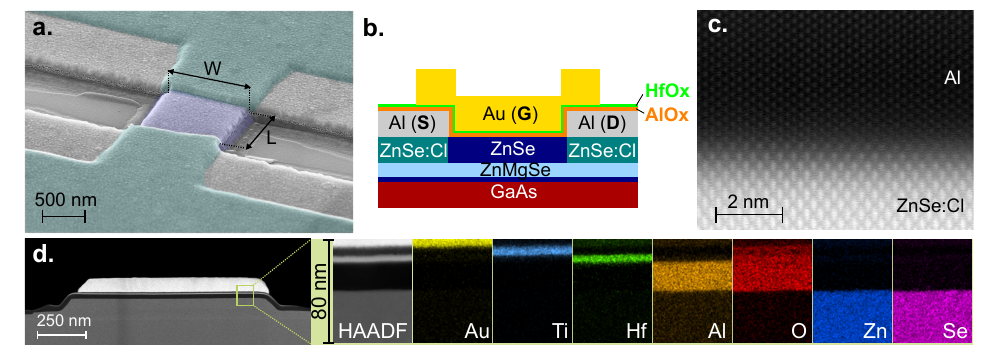}
    \caption{(a) SE SEM micrograph of a typical transistor device after selective removal of shadow walls. The false colors highlight the doped (green) and undoped (blue) regions of the underlying \ch{ZnSe}. (b) Illustration of the cross-section along the length of the channel with metal source (S), drain (D) and gate (G) contacts. HAADF-STEM images of (c) the source/drain contact region and (d) cross-section along the width of the channel region with elemental EDX maps highlighting the composition of the individual layers.}
    \label{fig:structure-2}    
\end{figure*}
A SEM micrograph of a typical FET channel width $W$ and length $L$ is shown in Fig.~\ref{fig:structure-2}~a. Note that the shadow walls were selectively removed for this image. The corresponding device cross-section is sketched in Fig.~\ref{fig:structure-2}~b, especially highlighting the \ch{ZnSe} channel (Ch) as well as the source (S), drain (D) and gate (G) contacts. The well-defined and abrupt shadowing of the material fluxes during growth is apparent in the different regions identified in the SEM image. In Fig.~\ref{fig:structure-2}~a the area in which \ch{ZnSe} was deposited is false-colored to identify the regions in which the underlying \ch{ZnSe} film is highly doped (green) and undoped (blue). Directly adjacent to the top and bottom edge of the walls the \ch{ZnSe} growth was suppressed, since either the \ch{Zn} or the \ch{Se} fluxes were shadowed by the wall in these regions. However, since \qty{20}{nm} of the \ch{ZnMgSe} layer were grown prior to shadow-wall fabrication, source and drain contacts remain electrically decoupled from the \ch{GaAs} substrate. Additional steps on the surface caused by the shadowing of the \ch{Al} can be clearly identified at either end of the undoped channel region. Furthermore, no cracks or gaps in the top \ch{Au} layer are visible, which confirms the uniform covering of the channel with the gate metal.\\
\\
Figure~\ref{fig:structure-2}~c shows a high-resolution high-angle annular dark-field scanning transmission electron microscopy (HAADF-STEM) image of the source/drain contact region. The \textit{in-situ} deposition of the \ch{Al} contact metal on top of \ch{ZnSe} leads to an atomically smooth interface. Furthermore, the \ch{Al} is grown single-crystalline on the \ch{ZnSe} surface promising excellent contact performance.\\
\\
A perpendicular HAADF-STEM cross-section of the channel is provided in Fig.~\ref{fig:structure-2}~d. The cross-section confirms the conformal covering of the entire channel with the ALD oxide. The chemical composition of the lamella is analyzed using spectrum imaging with STEM and energy-dispersive X-ray spectroscopy (EDX). The HAADF-STEM image and the corresponding elemental maps are shown as insets on the right side of Fig.~\ref{fig:structure-2}~d. All layers of the heterostructure, as illustrated in Fig.~\ref{fig:structure-2}~b, can be clearly distinguished. In particular, we observe a sharp chemical interface between the oxide stack and the \ch{ZnSe} itself, with no evidence of significant intermixing between the adjacent layers. Additional STEM results with EDX maps of the contact region and the \ch{ZnSe} heterostructure are provided in Section~S1 of the Supporting Information. Furthermore, we performed a detailed X-ray Diffraction (XRD) analysis to confirm the high crystal quality and the absence of strain relaxation (see Methods Section and Supporting Information Section~S2). This analysis additionally allows for precise extraction of the individual layer thicknesses.

%% file: chapters/electrical.tex
\begin{figure}[H]
    \centering
    \includegraphics[]{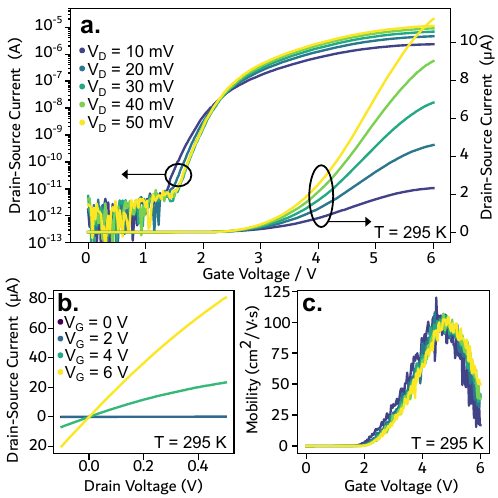}
    \caption{(a) Transfer characteristics of a \ch{ZnSe} FET (5 channels, $W=\qty{800}{nm}$, $L=\qty{1}{um}$) at room temperature plotted in semi-logarithmic and linear scale. (b) Output curves of the same \ch{ZnSe} transistor. (c) Gate-dependent field-effect mobility extracted from the transfer curves shown in (a).}
    \label{fig:RT}
\end{figure}
Figure~\ref{fig:RT}~a presents I-V measurements of a FET device consisting of 5 parallel $W=\qty{800}{nm}$ wide and $L=\qty{1}{\mu m}$ long channels. Consequently, the effective FET width is $W_\text{eff.}=5\cdot W = \qty{4}{\mu m}$. The measurements were performed in darkness at room temperature using a probe station (see Methods section). The drain-source current was recorded, while the gate voltage was swept at a constant drain voltage $V_{D}$. For all measurements, the source contact was grounded. For clarity, the data is plotted on both linear and semi-logarithmic scales. In the linear scale at a threshold voltage of  $V_{T}=$ \qty{3.5}{V}, a clear increase in drain-source current is observed. For gate voltages below the threshold $V_{T}$, the drain-source current increases non-linearly. In this region, a minimum sub-threshold swing of $SS_\text{min.}^{(295K)}= \qty{145}{mV\per dec}$ is extracted using $SS_\text{min.}^{(295K)}=\frac{\partial V_G}{\partial \log_{10}(I_D)}$ (See Supporting Information Section~S3). Comparable sub-threshold behaviour is also observed in conventional planar FET structures realized in other material systems, e.g. \ch{GaN}~\cite{Duan2026} or \ch{SiC}~\cite{Kang2023}. The increase of the sub-threshold swing above its theoretical limit of \qty{60}{mV\per dec} is indicative of progressively charged trap states located at the semiconductor oxide interface~\cite{Lyu1993}, which are common in material systems with less optimized oxide interfaces. The non-linear sub-threshold region is followed by a linear increase in drain-source current at moderate gate voltages. At higher gate voltages $V_G$ the drain current saturates. Note, that over the whole investigated gate voltage range the leakage to the gate is negligible and remains below \qty{3}{\nA} (Supporting Information Section~S3).\\
\\
To gain a deeper understanding of the transport behavior, we performed self-consistent Schrödinger–Poisson and drift–diffusion simulations (see Methods and Supporting Information, Section S4). The simulations reproduce the measured threshold voltage and show that the gradual opening of the conducting channel is strongly influenced by acceptor-like interface states at the \ch{ZnSe}/\ch{Al2O3} interface. These states are progressively charged as the gate voltage increases, delaying electron accumulation in the \ch{ZnSe} channel and resulting in the smooth onset of the drain-source current and  the observed degradation of the sub-threshold swing. The simulations predict an intrinsic low-field electron mobility of approximately $\mu_{0}\approx\qty{200}{cm\squared\per Vs}$ at room temperature. For gate voltages above approximately \qty{5}{V}, the simulations reveal strong downward bending of the \ch{ZnSe} conduction band, resulting in pronounced confinement of the electron gas at the \ch{ZnSe}/\ch{Al2O3} interface. The increased overlap of the electron ensemble with interface defects is expected to enhance interface scattering, leading to mobility degradation and the observed deviation of the drain-source current from its initial linear dependence on gate voltage.\\
\\
In addition to transfer characteristics, the output characteristics were measured by sweeping the drain voltage, $V_{D}$, while maintaining a constant gate voltage (Fig.~\ref{fig:RT}~b). At $V_G=\qty{0}{\V}$, the intrinisic \ch{ZnSe} channel is highly resistive and the drain-source current remains below the measurement limit of $\qty{e-12}{A}$. Such a highly resistive off-state is expected due to the wide band-gap of \ch{ZnSe} ($E_G\approx \qty{2.82}{eV}$~\cite{Wu2022}) and consequently low intrinsic carrier concentration. FETs with similarly highly resistive off-states are realized in other wide band-gap materials such as \ch{Ga2O3}~\cite{Song2023}, \ch{GaN}~\cite{Duan2026}  or \ch{SiC}~\cite{Kang2023}. Furthermore, the high resistance of the device at $V_G=\qty{0}{V}$ confirms the absence of \ch{Cl} dopants in the channel region and validates the SWE-approach for the selective doping of the \ch{ZnSe} film only in the contact region. At higher gate voltage $V_G$, electrons accumulate at the \ch{ZnSe}/oxide interface and  the channel resistance is decreased. The drain-source current reaches about \qty{80}{\uA} at $V_G=\qty{6}{\V}$ and $V_D=\qty{500}{\mV}$. Similar to other wide band-gap FETs~\cite{Song2023,Duan2026,Kang2023}, the investigated FET device shows a high $I_\text{on}$/$I_\text{off}$-ratio exceeding $>10^7$.\\
\\
The measured output curves are linear in the vicinity of $V_D=\qty{0}{V}$, indicating ohmic source and drain contacts to the channel. At higher drain voltages, the drain-source current exhibits a progressively weaker increase with $V_D$ and deviates from an ideal linear dependence. Self-consistent Schrödinger-Poisson and drift-diffusion simulations indicate that the increasing drain bias results in a non-uniform electron distribution along the channel (Supporting Information~S3). As the local channel potential increases towards the drain, the effective gate voltage is reduced, leading to weaker band bending, a lower electron density near the drain-side edge of the gate, and consequently a reduced charge carrier density. This effect, together with high-field degradation of the electron mobility, likely associated with enhanced optical phonon scattering, accounts for the observed sub-linear dependence of the drain-source current on the drain voltage $V_D$.\\
\\
In the linear region of the output curve, i.e., at small drain bias $V_D$, the drain-source current $I_D$ can be expressed as:
\begin{equation}\label{eq:ID-VG}
    I_D=\mu_{FE}\cdot C_{Ox}^\square\frac{W_\text{eff.}}{L}\left(V_{G}-V_T-\frac{V_D}{2}\right)\cdot V_D
\end{equation}
with the field-effect mobility $\mu_{FE}$, oxide capacitance per unit area $C_{Ox}^\square$ and threshold voltage $V_T$.~\cite{Waser2003} The oxide capacitance is extracted separately from capacitor structures using the same ALD oxide stack (Supporting Information Section~S5). In a common approach, the field-effect mobility $\mu_{FE}$ is obtained from the transconductance $g_m=\frac{\partial I_D}{\partial V_G}$ using:
\begin{equation}\label{eq:mobility}
    \mu_{FE}=\frac{g_m}{C_{Ox}^\square\cdot V_{DS}}\cdot\frac{L}{W_\text{eff.}}
\end{equation}
The extraction of the field-effect mobility from the transconductance assumes the case of an ideal MOSFET with perfect ohmic contacts to the semiconductor channel. This might lead to over- or underestimation of the true field-effect mobility in real devices~\cite{Liu2017}. However, this form of mobility extraction neither requires precise determination of the threshold voltage $V_T$ nor specific assumptions about contributing scattering mechanisms.\\
\\
The pronounced maximum in transconductance leads to a peak in the field-effect mobility as a function of gate voltage as shown in Fig.~\ref{fig:RT}~c. The maximum extracted field-effect mobility is around $\mu_{FE}^{(295K)}=\qty{101}{cm\squared\per Vs}$ independent of drain bias $V_{D}$. The reduction of the measured field-effect mobility $\mu_{FE}$ compared to the simulated mobility $\mu_0$ is attributed to additional scattering mechanisms, including interface-related scattering and the influence of the strong vertical electric field at the \ch{ZnSe}/\ch{Al2O3} interface. Further evidence for additional scattering related to carrier confinement at the semiconductor/oxide interface is the reduction in mobility compared to the Hall mobility measured in bulk \ch{ZnSe} samples (Supporting Information Section~S6). Therefore, similar to \ch{Ge}-based systems~\cite{Saraswat2006,Gupta2013}, careful optimization of the gate oxide stack and surface preparation prior to oxide deposition specifically for the \ch{ZnSe} system promises strong reduction of the number of interface traps and significant improvement of the performance of \ch{ZnSe}-based devices. Furthermore, as for \ch{Ge}-based systems~\cite{Lodari2021,Kaul2025}, we expect the influence of interface traps to be strongly reduced in a quantum well heterostructure, where the conducting channel is not located directly at the oxide interface.\\
\\
\begin{figure}
    \centering
    \includegraphics[]{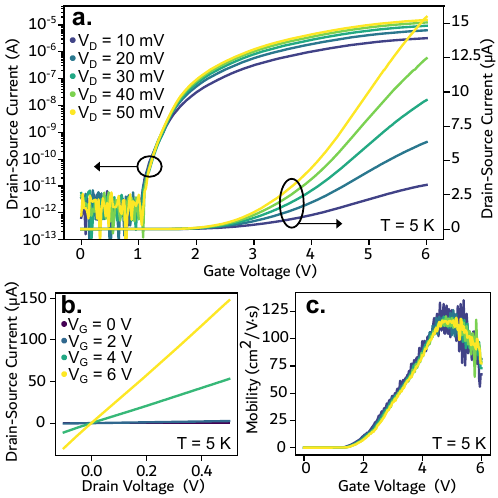}
    \caption{(a) Transfer characteristics of the same \ch{ZnSe} transistor (5 channels, $W=\qty{800}{nm}$, $L=\qty{1}{um}$) presented in Fig.~\ref{fig:RT} but at $T=\qty{5}{K}$ plotted in semi-logarithmic and linear scale and (b) corresponding output curves. (c) Gate-dependent field-effect mobility extracted from the transfer curves shown in (a).}
    \label{fig:5K_TLM800x5}
\end{figure}
To assess the potential of \ch{ZnSe} heterostructures for spin-qubit applications, we additionally investigated the low-temperature transport behavior at \qty{5}{K} (see Fig.~\ref{fig:5K_TLM800x5}). Compared with room-temperature operation, the transfer characteristic (Fig.~\ref{fig:5K_TLM800x5}~a) exhibits a reduced minimum sub-threshold slope of $SS_\text{min.}^{(5K)}= \qty{73}{mV/dec}$, while maintaining a highly resistive off-state. The increase in drain-source current in the on-state with decreasing temperature is attributed to enhanced carrier transport due to reduced phonon scattering~\cite{Ruda1986}. This is also apparent in the output curves (Fig.~\ref{fig:5K_TLM800x5}~b), which remain linear over the whole investigated voltage range. Quantitatively the enhanced carrier transport is reflected in the maximum field-effect mobility (Fig.~\ref{fig:5K_TLM800x5}~c), which is noticeably higher than at room temperature reaching $\mu_{FE}=116\frac{\text{cm}^2}{V\cdot s}$ at \qty{5}{K}. \\
\\
The overall improved transport characteristics at $T=\qty{5}{K}$ indicate that transport is dominated by the channel rather than being contact-limited and is characteristic for devices with low-resistance ohmic source/drain contacts~\cite{Shen2021}. Especially the linear output characteristic (Fig.~\ref{fig:5K_TLM800x5}~b) confirms that the \ch{Al} source and drain contacts do not freeze out and maintain their ohmic behavior. Furthermore, the abscence of additional barriers at \qty{5}{K} indicates sufficient overlap of the highly-doped source/drain regions and the channel region, where the carrier density is gate-controlled. This observation in particular highlights that the doping concentration within the \ch{ZnSe} layer can be precisely controlled using the SWE-technique, which allows for the formation of local ohmic contacts to a gated \ch{ZnSe} heterostructure.\\
\\
\begin{figure*}
    \centering
    \includegraphics[]{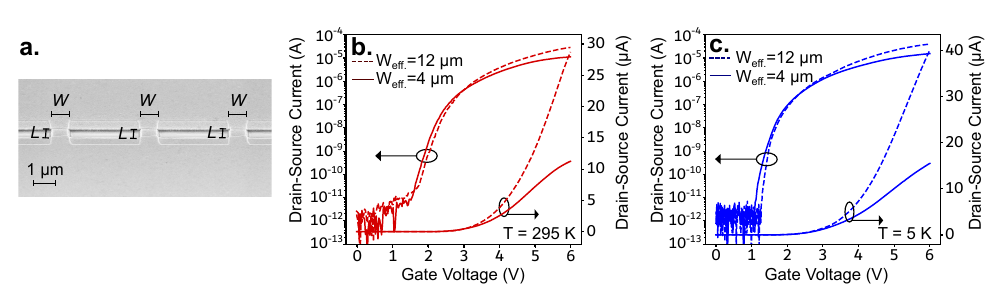}
    \caption{(a) SEM micrograph showing three parallel transistor channels of equal width $W$ and length $L$. Transfer curves of transistors with $W_\text{eff.}=\qty{4}{\um}$ and $W_\text{eff.}=\qty{12}{\um}$ at (b) \qty{295}{K} and at (c) \qty{5}{K}}.
    \label{fig:compareN}
\end{figure*}
As illustrated by the SEM image in Fig.~\ref{fig:compareN}~a, the investigated FET devices consist of multiple parallel channels. We electrically characterize the performance of an additional \ch{ZnSe} FET on the same sample that consists of 15 parallel $W=\qty{800}{nm}$ wide and $L=\qty{1}{\um}$ long channels. The effective channel width for this FET is therefore $W_\text{eff.}=15\cdot W = \qty{12}{\um}$. In Fig.~\ref{fig:compareN}~b transfer curves of both the 5-channel FET ($W_\text{eff.}=\qty{4}{\um}$, solid line) and the 15-channel FET ($W_\text{eff.}=\qty{12}{\um}$, dashed line) at  $T$ = \qty{295}{K} are shown. Sub-threshold swing, maximum transconductance and corresponding field-effect mobility extracted for both transistors are summarized in Tab.~\ref{tab:compare}. Both devices exhibit similar behaviour, most notably a highly resistive off-state below the measurement limit, a high on-off ratio of $>10^7$ and a similar subthreshold swing. A significant difference is the three-fold increase in drain-source current and in the maximum transconductance for the 15-channel devices, due to the increase in effective channel width $W_\text{eff.}$ (see Eq.~\ref{eq:ID-VG}). As a result, the field-effect mobility extracted from the maximum transconductance is similar (see Eq.~\ref{eq:mobility}), indicating comparable properties of the individual \ch{ZnSe} channels in both devices.\\
\\
The trapping and de-trapping of carriers at defects at the oxide/semiconductor interface as well as defects within the oxide layer cause a characteristic clock-wise hysteresis of the gate sweep (Supporting Information S3). The hysteresis $\Delta V$ is extracted in the sub-threshold region at a constant current of $I_D=\qty{10}{nA}$ and summarized in Tab.~\ref{tab:compare}. The hysteresis is lower for the 15-channel device, which is consistent with the increased mobility for this device compared to the 5-channel FET.\\
\begin{table}
    \centering
    \begin{tabular}{|c|c|c|c|c|}\hline
         &\multicolumn{2}{c|}{$W_\text{eff}=$\qty{4}{\um}}& \multicolumn{2}{c|}{$W_\text{eff}=$\qty{12}{\um}}\\\cline{2-5}
         &\qty{295}{K}&\qty{5}{K}&\qty{295}{K}&\qty{5}{K}\\\hline
         $SS_\text{min.}$ (mV/dec)&145&73& 167&49\\\hline
         $g_{m,\text{max.}}$ (\textmu A/V)&5.3&6.1&16.9&19.5\\\hline
         $\mu_{FE}$ ($\text{cm}^2/Vs$)&101&116&107&123\\\hline
         $\Delta V$ (mV)&420&42&295&49\\\hline
    \end{tabular}
    \caption{Summary of parameters extracted from transfer curves of two MOSFETs with 5-channels ($W_\text{eff}=$\qty{4}{\um}) and 15-channels ($W_\text{eff}=$\qty{12}{\um}).}
    \label{tab:compare}
\end{table}
\\
Transfer curves for both FETs at $T=\qty{5}{K}$ are shown in Fig.~\ref{fig:compareN}~c. For both devices, the drain-source current and the maximum transconductance increase compared to the room temperature measurement. The previously-described scaling by a factor of three between the two devices persists at \qty{5}{K} and is consistent with the increased effective channel width $W_\text{eff.}$ for the 15-channel FET.
The reliable scaling between the two devices additionally supports the formation of high quality ohmic contacts and a sufficient overlap of the source/drain regions with the gate-controlled channel region for the individual channels in both devices. In the supporting information~S3, we provide output curves of the 15-channel FET confirming the ohmic contact behaviour as well as long-term current stability exceeding \qty{15}{min}. The latter is particularly relevant for future spin-qubit applications, in which current stability directly impacts the read-out fidelity~\cite{Elzerman2004}. The gate hysteresis for both devices is strongly suppressed at \qty{5}{K}, which points to a freezing out of the acceptor-like interface states~\cite{Park2021}. The control of the gate-loop hysteresis is essential for the reliable tuning of spin qubit devices~\cite{Massai2024}. The consistent scaling of FETs consisting of multiple parallel channels underscores the robustness of the SWE-based fabrication approach, while the demonstrated current stability and negligible gate hysteresis confirms that the investigated \ch{ZnSe} heterostructure offers the electrostatic control required for future gate-defined spin-qubit devices.

%% file: chapters/conclusion.tex
In this work, we combined a novel technique for MBE growth, which utilizes the selective shadowing of directional material fluxes at pre-structured shadow walls, with the formation of superior ohmic contacts via \textit{in-situ} deposition of the \ch{Al} contact metal to realize FETs in \ch{ZnSe}. A detailed structural and chemical analysis confirmed the selective epitaxial growth using the SWE technique and the successful realization of the intended heterostructure. In particular, the atomically smooth interface of the single-crystalline \ch{Al} contact metal to the \ch{ZnSe:Cl} highlights the potential of the fabrication method for the realization of low-resistance local ohmic contacts. 
The final devices unambiguously demonstrate the expected FET behaviour such as modulation of the channel resistance with applied gate voltage and linear output characteristics. However, as commonly observed in other material platforms, device performance remains constraint by defects located at the \ch{ZnSe}/oxide interface. Therefore, careful optimization of the \ch{ZnSe}/oxide interface as well as integration of the \ch{ZnSe} channel into a quantum well heterostructure are promising approaches to mitigate the effect of these interface traps. Transport characterization at room temperature and at $T=\qty{5}{K}$ confirms that local low-resistance ohmic contacts to the \ch{ZnSe} channel, as well as sufficient overlap of the highly-doped contacts with the gate-controlled region of the undoped \ch{ZnSe} channel are realized using the SWE-based fabrication technique. Furthermore, we observe an overall improved device performance at \qty{5}{K}, in particular a reduction of the sub-threshold swing, a decrease in gate hysteresis and an increase of the field-effect mobility. These observation highlights that the developed fabrication method is capable of producing devices which are not limited by contact performance and allow for electrostatic control of the carrier concentration in the undoped \ch{ZnSe}-channel at \qty{5}{K}. Furthermore, the consistent scaling of FETs consisting of multiple parallel channels confirms the robustness of the developed fabrication approach. The development of a reliable fabrication scheme for \ch{ZnSe}-based heterostructures that operate at cryogenic temperatures provides a toolbox for the future exploration of \ch{ZnSe} as a material for spin-qubit applications.

%% file: chapters/methods.tex
\subsection{MBE growth}

All MBE growth was performed in an MBE cluster system, interconnected by UHV transfer lines, allowing for the monolithic combination of different materials. The substrate of choice was epi-ready \ch{GaAs} (001) substrate, which offers a small lattice mismatch of \qty{0.25}{\percent} to ZnSe. Before shadow wall fabrication, a \ch{ZnMgSe} buffer was prepared in order to minimize leakage into the \ch{GaAs} substrate. A \qty{300}{\nm} thick \ch{GaAs} layer was grown within a dedicated MBE chamber for the epitaxy of III/V semiconductors at a substrate temperature of $T_{sub}$ = \qty{600}{\celsius}. This provides a pristine interface for the subsequent II/VI growth~\cite{vondenDriesch2024}. The substrate was then transferred under UHV conditions into another MBE chamber designed for growth of II/VI selenides. Growth was initiated with an approximately \qty{10}{\nm} \ch{ZnSe} layer to achieve an abrupt interface to the \ch{GaAs} buffer. This was followed by a \qty{20}{\nm} thick \ch{ZnMgSe} film, serving as the actual buffer layer for the subsequent FET devices. All II/VI layers are grown under \ch{Se}-rich conditions at a substrate temperature of approximately \qty{290}{\celsius}. Finally, this II/VI buffer was \textit{in-situ} passivated with \qty{20}{\nm} of \ch{AlO_x} evaporated via an electron-beam inside another MBE chamber of the cluster system. This passivation layer prevents oxidation of the buffer and protects the substrate from further degradation during the subsequent shadow wall fabrication.\\
\\
Following the final etching step of the shadow wall fabrication, the sample was immediately reintroduced into the MBE cluster loadlock to minimize the oxidation of the \ch{ZnMgSe} buffer. The sample was then baked within the loadlock at \qty{200}{\degree C} for 45\,min to remove any volatile adsorbates or water vapor. Additionally, consistent with the epitaxy of \ch{ZnSe} on III/V substrates\cite{Zhang2021}, an \textit{in-situ} cleaning step using atomic hydrogen was used to remove carbon and oxide residuals from the \ch{ZnMgSe} surface~\cite{CONTACT}. Subsequently, the substrate was transferred under UHV conditions into the II/VI MBE chamber. Here, the FET heterostructure consisting of additional \qty{20}{\nm} of \ch{ZnMgSe} and \qty{50}{\nm} of highly-doped \ch{ZnSe:Cl} was grown under identical growth conditions as the previous \ch{ZnMgSe} buffer. It should be noted that precise sample orientation during this step is essential, as the alignment of the material fluxes relative to the shadow walls is critical for the realization of such transistors via SWE.

\subsection{Shadow Wall and Gate Contact Fabrication}
All cleanroom processing was performed in the Helmholtz Nanofacility at the Forschungszentrum Juelich GmbH~\cite{Albrecht2017}. As a first step for shadow wall fabrication, a stack of \qty{150}{\nm} \ch{SiO_x} and \qty{1.8}{\um} \ch{Si} was deposited on the AlOx-passivated \ch{ZnMgSe}-buffer. A \ch{BaF2} hard-mask was defined on top of the \ch{Si}-based wall-stack using electron beam lithography and a lift-off process.
The \ch{Si} layer was subsequently etched using inductively coupled plasma (ICP) reactive ion etching with a combination of \qty{100}{sccm} \ch{SF6}, \qty{5}{sccm} \ch{CH4} and \qty{10}{sccm} \ch{Ar}. The etching selectivity to the \ch{BaF2} hardmask is excellent with the \ch{Si} etch-rate of about \qty{60}{\nm\per\minute} compared to the \ch{BaF2} etch-rate of only \qty{2}{\nm\per\minute}. Finally, the remaining \ch{SiO_x} and \ch{AlO_x} layers were etched through in a single wet-etching step using diluted hydrofluoric acid (HF \qty{1}{\%}), which simultaneously formed an undercut beneath the Si layer to prevent electrical shorts.. This step was performed within \qty{10}{\minute} of re-introducing the sample into the UHV system to minimize oxidation of the \ch{ZnMgSe} buffer.\\
\\
Following the MBE growth of the heterostructure, the deposition of the source/drain contacts and the ALD passivation, the gate contacts were fabricated. For this purpose, the gate metal stack - comprising a \qty{10}{\nm} Ti adhesion layer and a \qty{75}{\nm} Au layer - was patterned via a lift-off process using an optical maskless lithography system (MLA100, Heidelberg Instruments).
\subsection{Structural Characterization}
The high crystal quality of the MBE grown heterostructures was confirmed using X-ray diffraction (XRD) measurements. We measured $\theta$-$2\theta$ scans and rocking curves around the (004) reflex, as well as a reciprocal space maps in a Rigaku SmartLab XRD system (see Supporting information, Section S2). Additional X-ray reflectometry (XRR) measurements allow for precise determination of the thicknesses of the \ch{Al} contact metal and the ALD passivation layer. Furthermore, electron probe aberration-corrected STEM systems (ThermoFisher Scientific Titan ChemiSTEM 80-200 and Spectra) operated at 200~kV and 300~kV acceleration voltages were used to confirm the high-quality of the ohmic contacts and atomically smooth interfaces throughout the entire heterostructure. STEM/EDX spectrum imaging reveals the chemical distribution and composition of the layers. The electron transparent specimens for STEM were prepared perpendicular and parallel to the channel using a Xe plasma focused ion beam (FIB) sputtering in a dual-beam SEM system following the conventional lift-out method. 
\subsection{Electrical Characterization}
Electrical characterization of the transistors at room temperature was performed using a Keithley 4200 coupled to a needle probestation in the Helmholtz Nano Facility (HNF)~\cite{Albrecht2017}. This system also allows for dielectric characterization via two-point CV measurements. Temperature dependent measurements were performed using a PLC50 probe system with an Agilent E5270B parameter analyzer, a four point needle probe station that allows for measurements down to a base temperature of \qty{5}{K}.
\subsection{Simulation}
The electrostatic and transport properties of the \ch{ZnSe} FET were analyzed using self-consistent numerical simulations performed with nextnano++. The Schrödinger-Poisson equations were solved to obtain the gate-induced band bending and electron distribution in the ZnSe channel, while the drift-diffusion model was used to calculate the current-voltage characteristics. Interface charge states at the \ch{ZnSe}/\ch{Al2O3} interface were included to account for their influence on channel formation and threshold voltage. The material parameters, band offsets, and simulation details are provided in the Supporting Information.

%% file: bibliography.bib
@article{vondenDriesch2024,
   author = {Nils von den Driesch and Yurii Kutovyi and Felix Khamphasithivong and Anja Zass and Lars R. Schreiber and Alexander Pawlis},
   doi = {10.1021/acsaelm.4c01104},
   issn = {26376113},
   journal = {ACS Applied Electronic Materials},
   publisher = {American Chemical Society},
   title = {Shadow Wall Epitaxy of Compound Semiconductors toward All in Situ Fabrication of Quantum Devices},
   year = {2024}
}

@article{Jansen2020,
   author = {Johanna Janßen and Felix Hartz and Till Huckemann and Christian Kamphausen and Malte Neul and Lars R. Schreiber and Alexander Pawlis},
   doi = {10.1021/acsaelm.9b00824},
   issn = {26376113},
   issue = {4},
   journal = {ACS Applied Electronic Materials},
   month = {4},
   pages = {898-905},
   publisher = {American Chemical Society},
   title = {Low-Temperature Ohmic Contacts to n-ZnSe for all-Electrical Quantum Devices},
   volume = {2},
   year = {2020}
}

@article{Albrecht2017,
  title = {{HNF} - {H}elmholtz {N}ano {F}acility},
  author = {Wolfgang Albrecht and Juergen Moers and Bernd Hermanns},
  year = {2017},
  publisher = {Forschungszentrum Julich, Zentralbibliothek},
  journal = {Journal of large-scale research facilities JLSRF},
  volume = {3},
  pages = {A112},
  doi = {10.17815/jlsrf-3-158},
  url = {https://doi.org/10.17815/jlsrf-3-158}
}

@article{Kirstein2021,
   author = {Erik Kirstein and Evgeny A. Zhukov and Dmitry S. Smirnov and Vitalie Nedelea and Phillip Greve and Ina V. Kalitukha and Viktor F. Sapega and Alexander Pawlis and Dmitri R. Yakovlev and Manfred Bayer and Alex Greilich},
   doi = {10.1038/s43246-021-00198-z},
   issn = {26624443},
   issue = {1},
   journal = {Communications Materials},
   month = {12},
   publisher = {Springer Nature},
   title = {Extended spin coherence of the zinc-vacancy centers in ZnSe with fast optical access},
   volume = {2},
   year = {2021}
}

@article{Pawlis2019,
   author = {Alexander Pawlis and Gregor Mussler and Christoph Krause and Benjamin Bennemann and Uwe Breuer and Detlev Grutzmacher},
   doi = {10.1021/acsaelm.8b00006},
   issn = {26376113},
   issue = {1},
   journal = {ACS Applied Electronic Materials},
   month = {1},
   pages = {44-50},
   publisher = {American Chemical Society},
   title = {MBE Growth and Optical Properties of Isotopically Purified ZnSe Heterostructures},
   volume = {1},
   year = {2019}
}

@article{Burkard2023,
   author = {Guido Burkard and Thaddeus D. Ladd and Andrew Pan and John M. Nichol and Jason R. Petta},
   doi = {10.1103/RevModPhys.95.025003},
   issn = {15390756},
   issue = {2},
   journal = {Reviews of Modern Physics},
   month = {4},
   publisher = {American Physical Society},
   title = {Semiconductor spin qubits},
   volume = {95},
   year = {2023}
}

@article{Zhang2021,
   author = {Chaomin Zhang and Kirstin Alberi and Christiana Honsberg and Kwangwook Park},
   doi = {10.1016/j.apsusc.2021.149245},
   issn = {01694332},
   journal = {Applied Surface Science},
   month = {5},
   publisher = {Elsevier B.V.},
   title = {Investigation of GaAs surface treatments for ZnSe growth by molecular beam epitaxy without a buffer layer},
   volume = {549},
   year = {2021}
}

@article{Dreifus1990,
   author = {D. L. Dreifus and B. P. Sneed and J. Ren and J. W. Cook and J. F. Schetzina and R. M. Kolbas},
   doi = {10.1063/1.104079},
   issn = {00036951},
   issue = {16},
   journal = {Applied Physics Letters},
   pages = {1663-1665},
   title = {ZnSe field-effect transistors},
   volume = {57},
   year = {1990}
}

@article{Liu2017,
   author = {Chuan Liu and Gongtan Li and Riccardo Di Pietro and Jie Huang and Yong Young Noh and Xuying Liu and Takeo Minari},
   doi = {10.1103/PhysRevApplied.8.034020},
   issn = {23317019},
   issue = {3},
   journal = {Physical Review Applied},
   month = {9},
   publisher = {American Physical Society},
   title = {Device Physics of Contact Issues for the Overestimation and Underestimation of Carrier Mobility in Field-Effect Transistors},
   volume = {8},
   year = {2017}
}

@article{Gunshor1988,
   author = {Robert L. Gunshor and Leslie A. Kolodziejski},
   doi = {10.1109/3.7104},
   issue = {8},
   journal = {IEEE Journal of Quantum Electronics},
   month = {8},
   pages = {1744-1757},
   title = {Recent Advances in the Molecular Beam Epitaxy of the Wide-Bandgap Semiconductor ZnSe and Its Superlattices},
   volume = {24},
   year = {1988}
}

@article{Shen2021,
   author = {Pin Chun Shen and Cong Su and Yuxuan Lin and Ang Sheng Chou and Chao Ching Cheng and Ji Hoon Park and Ming Hui Chiu and Ang Yu Lu and Hao Ling Tang and Mohammad Mahdi Tavakoli and Gregory Pitner and Xiang Ji and Zhengyang Cai and Nannan Mao and Jiangtao Wang and Vincent Tung and Ju Li and Jeffrey Bokor and Alex Zettl and Chih I. Wu and Tomás Palacios and Lain Jong Li and Jing Kong},
   doi = {10.1038/s41586-021-03472-9},
   issn = {14764687},
   issue = {7858},
   journal = {Nature},
   month = {5},
   pages = {211-217},
   pmid = {33981050},
   publisher = {Nature Research},
   title = {Ultralow contact resistance between semimetal and monolayer semiconductors},
   volume = {593},
   year = {2021}
}

@article{Riley1996,
   author = {J Riley and D Wolfframm and D Westwood and A Evans},
   journal = {Journal of Crystal Growth},
   pages = {193-200},
   title = {Studies in the growth of ZnSe on GaAs(001)},
   volume = {160},
   year = {1996}
}

@article{Schallenberg2003,
    title = {{Projective techniques for the growth of compound semiconductor nanostructures}},
    year = {2003},
    journal = {Physica Status Solidi (A) Applied Research},
    author = {Schallenberg, Timo and Schumacher, C. and Molenkamp, Laurens W.},
    number = {1 SPEC},
    pages = {232--237},
    volume = {195},
    doi = {10.1002/pssa.200306291},
    issn = {00318965}
}

@article{Continenza1988,
   author = {A Continenza and S Massidda and A J Freeman},
   doi = {10.1103/PhysRevB.38.12996},
   issue = {18},
   journal = {Physical Review B},
   month = {12},
   pages = {12996-13001},
   title = {Structural and electronic properties of bulk ZnSe},
   volume = {38},
   url = {10.1103/PhysRevB.38.12996},
   year = {1988}
}

@article{Ruda1986,
   author = {H. E. Ruda},
   doi = {10.1063/1.336509},
   issn = {00218979},
   issue = {4},
   journal = {Journal of Applied Physics},
   pages = {1220-1231},
   title = {A theoretical analysis of electron transport in ZnSe},
   volume = {59},
   year = {1986}
}

@article{Kuttler1996,
   author = {M. Kuttler and M. Strassburg and V. Türck and R. Heitz and U. W. Pohl and D. Bimberg and E. Kurtz and G. Landwehr and D. Hommel},
   doi = {10.1063/1.117546},
   issn = {00036951},
   issue = {18},
   journal = {Applied Physics Letters},
   month = {10},
   pages = {2647-2649},
   publisher = {American Institute of Physics Inc.},
   title = {Laterally structured ZnCdSe/ZnSe superlattices by diffusion induced disordering},
   volume = {69},
   year = {1996}
}

@article{Wu2022,
   author = {Yifeng Wu and Kelsey J. Mirrielees and Douglas L. Irving},
   doi = {10.1063/5.0092736},
   issn = {00036951},
   issue = {23},
   journal = {Applied Physics Letters},
   month = {6},
   publisher = {American Institute of Physics Inc.},
   title = {On native point defects in ZnSe},
   volume = {120},
   year = {2022}
}

@article{Xue2022,
  author="Xue, Xiao and Russ, Maximilian and Samkharadze, Nodar and Undseth, Brennan and Sammak, Amir and Scappucci, Giordano and Vandersypen, Lieven M. K.",
  title="Quantum logic with spin qubits crossing the surface code threshold",
  journal="Nature",
  volume="601",
  number="7893",
  pages="343",
  publisher="Nature Publishing Group",
  year="2022",
  doi="10.1038/s41586-021-04273-w",
}

@article{Noiri2022,
	title = {Fast universal quantum gate above the fault-tolerance threshold in silicon},
	volume = {601},
	issn = {1476-4687},
	doi = {10.1038/s41586-021-04182-y},
	number = {7893},
	journal = {Nature},
	author = {Noiri, Akito and Takeda, Kenta and Nakajima, Takashi and Kobayashi, Takashi and Sammak, Amir and Scappucci, Giordano and Tarucha, Seigo},
	month = jan,
	year = {2022},
	pages = {338},
}

@article{Mills2022,
  title   = {Two-qubit silicon quantum processor with operation fidelity exceeding 99\%},
  author  = {Mills, A. R. and Guinn, C. R. and Gullans, M. J. and Sigillito, A. J. and Feldman, M. M. and Nielsen, E. and Petta, J. R.},
  journal = {Sci. Adv.},
  volume  = {8},
  pages   = {5130},
  year    = {2022},
  doi     = {10.1126/sciadv.abn5130}
}

@article{Wu2025,
      title={Simultaneous High-Fidelity Single-Qubit Gates in a Spin Qubit Array}, 
      author={Yi-Hsien Wu and Leon C. Camenzind and Patrick Bütler and Ik Kyeong Jin and Akito Noiri and Kenta Takeda and Takashi Nakajima and Takashi Kobayashi and Giordano Scappucci and Hsi-Sheng Goan and Seigo Tarucha},
      year={2025},
      journal={arXiv preprint 2507.11918}
}

@article{defuentes2025,
      title={Running a six-qubit quantum circuit on a silicon spin qubit array}, 
      author={I. Fernández de Fuentes and E. Raymenants and B. Undseth and O. Pietx-Casas and S. Philips M. Mądzik and S. L. de Snoo and S. V. Amitonov and L. Tryputen and A. T. Schmitz and A. Y. Matsuura and G. Scappucci and L. M. K. Vandersypen},
      year={2025},
      journal={arXiv preprint 2505.19200},
      preprint={Preprint at https://arxiv.org/abs/2505.19200},
      archivePrefix={arXiv},
      primaryClass={quant-ph},
}

@article{HRL2026,
author = {\text{Members of HRL Quantum Team and Collaborators}},
title={A digitally controlled silicon quantum processing unit},
journal={Nature},
year={2026},
month={Jul},
day={01},
volume={655},
number={8125},
pages={1154-1159},
issn={1476-4687},
doi={10.1038/s41586-026-10754-7},
url={https://doi.org/10.1038/s41586-026-10754-7}
}

@article{Boter2022,
  title = {Spiderweb Array: A Sparse Spin-Qubit Array},
  author = {Boter, Jelmer M. and Dehollain, Juan P. and van Dijk, Jeroen P.G. and Xu, Yuanxing and Hensgens, Toivo and Versluis, Richard and Naus, Henricus W.L. and Clarke, James S. and Veldhorst, Menno and Sebastiano, Fabio and Vandersypen, Lieven M.K.},
  journal = {Phys. Rev. Appl.},
  volume = {18},
  issue = {2},
  pages = {024053},
  numpages = {20},
  year = {2022},
  month = {Aug},
  publisher = {American Physical Society},
  doi = {10.1103/PhysRevApplied.18.024053},
}

@article{kunne2024spinbus,
  title={The SpinBus architecture for scaling spin qubits with electron shuttling},
  author={K{\"u}nne, Matthias and Willmes, Alexander and Oberl{\"a}nder, Max and Gorjaew, Christian and Teske, Julian D and Bhardwaj, Harsh and Beer, Max and Kammerloher, Eugen and Otten, Ren{\'e} and Seidler, Inga and others},
  journal={Nature Communications},
  volume={15},
  number={1},
  pages={4977},
  year={2024},
  publisher={Nature Publishing Group UK London}
}

@article{George2025,
	title = {12-{Spin}-{Qubit} {Arrays} {Fabricated} on a 300 mm {Semiconductor} {Manufacturing} {Line}},
	volume = {25},
	issn = {1530-6984},
	number = {2},
	journal = {Nano Letters},
	author = {George, Hubert C. and Mądzik, Mateusz T. and Henry, Eric M. and Wagner, Andrew J. and Islam, Mohammad M. and Borjans, Felix and Connors, Elliot J. and Corrigan, J. and Curry, Matthew and Harper, Michael K. and Keith, Daniel and Lampert, Lester and Luthi, Florian and Mohiyaddin, Fahd A. and Murcia, Sandra and Nair, Rohit and Nahm, Rambert and Nethwewala, Aditi and Neyens, Samuel and Patra, Bishnu and Raharjo, Roy D. and Rogan, Carly and Savytskyy, Rostyslav and Watson, Thomas F. and Ziegler, Josh and Zietz, Otto K. and Pellerano, Stefano and Pillarisetty, Ravi and Bishop, Nathaniel C. and Bojarski, Stephanie A. and Roberts, Jeanette and Clarke, James S.},
	month = jan,
	year = {2025},
	pages = {793--799},
}

@article{Struck2020,
  author="Struck, Tom and Hollmann, Arne and Schauer, Floyd and Fedorets, Olexiy and Schmidbauer, Andreas and Sawano, Kentarou and Riemann, Helge and Abrosimov, Nikolay V. and Cywinski, Lukasz and Bougeard, Dominique and Schreiber, Lars R.",
  title="Low-frequency spin qubit energy splitting noise in highly purified $^{28}${Si/SiGe}",
  journal="npj Quantum Inf.",
  volume="6",
  number="69",
  pages="2056",
  publisher="Nature Publishing Group",
  year="2020",
}

@article{Volmer2026,
    title={Impact of the local valley splitting on the coherence of conveyor-belt spin shuttling in $^{28}$\text{Si/SiGe}}, 
    author={Mats Volmer and Tom Struck and Arnau Sala and Jhih-Sian Tu and Stefan Trellenkamp and Davide Degli Esposti and Giordano Scappucci and Łukasz Cywiński and Hendrik Bluhm and Lars R. Schreiber},
    journal   = {Nature Communications},
    volume    = {17},
    pages     = {5448},
    year      = {2026},
}

@article{desmet2024,
	title = {High-fidelity single-spin shuttling in silicon},
	volume = {20},
	issn = {1748-3395},
	number = {7},
	journal = {Nat. Nanotechnol.},
	author = {De Smet, Maxim and Matsumoto, Yuta and Zwerver, Anne-Marije J. and Tryputen, Larysa and de Snoo, Sander L. and Amitonov, Sergey V. and Katiraee-Far, Sam R. and Sammak, Amir and Samkharadze, Nodar and Guel, Oender and Wasserman, Rick N. M. and Greplová, Eliška and Rimbach-Russ, Maximilian and Scappucci, Giordano and Vandersypen, Lieven M. K.},
	month = jul,
	year = {2025},
	pages = {866--872},
}

@article{Matsumoto2026,
  author    = {Y. Matsumoto and M. De Smet and L. Tryputen and S. L. de Snoo and S. V. Amitonov and A. Sammak and M. Rimbach-Russ and G. Scappucci and L. M. K. Vandersypen},
  title     = {Two-qubit logic and teleportation with mobile spin qubits in silicon},
  journal   = {Nature},
  volume    = {650},
  pages     = {56--61},
  year      = {2026}
}

@article{Beer2026,
      title={Conveyor-mode electron shuttling through a T-junction in Si/SiGe}, 
      author={Max Beer and Ran Xue and Lennart Deda and Stefan Trellenkamp and Jhih-Sian Tu and Paul Surrey and Inga Seidler and Hendrik Bluhm and Lars R. Schreiber},
      year={2026},
      journal={arXiv preprint 2601.03942}
}

@article{Xue2024,
  author    = {Ran Xue and Max Beer and Inga Seidler and Simon Sebastian Humpohl and Jhih-Sian Tu and Stefan Trellenkamp and Tom Struck and Hendrik Bluhm and Lars R. Schreiber},
  title     = {Si/\text{SiGe} QuBus for single electron information-processing devices with memory and micron-scale connectivity function},
  journal   = {Nature Communications},
  volume    = {15},
  pages     = {2296},
  year      = {2024},
}

@article{klos_atomistic_2024,
	title = {Atomistic Compositional Details and Their Importance for Spin Qubits in Isotope‐Purified Silicon Quantum Wells},
	volume = {11},
	number = {42},
	journal = {Adv. Sci.},
	author = {Klos, Jan and Tröger, Jan and Keutgen, Jens and Losert, Merritt P. and Abrosimov, Nikolay V. and Knoch, Joachim and Bracht, Hartmut and Coppersmith, Susan N. and Friesen, Mark and Cojocaru‐Mir{\'e}din, Oana and Schreiber, Lars R. and Bougeard, Dominique},
	month = nov,
	year = {2024},
	pages = {2407442},
}

@article{Muster2025Shuttling,
  author    = {P. Muster and
               W. Langheinrich and
               T. Huckemann and
               S. Pregl and
               V. Brackmann and
               M. Friedrich and
               F. Reichmann and
               N. D. Komeri\v{c}ki and
               L. R. Schreiber and
               J. Bluhm},
  title     = {High-Fidelity Single-Electron Shuttling in Industrially Fabricated Spin Qubit Devices},
  booktitle = {2025 IEEE International Electron Devices Meeting (IEDM)},
  year      = {2025},
  pages     = {1--4},
  publisher = {IEEE},
  doi       = {10.1109/IEDM50572.2025.11353490},
  url_none  = {https://ieeexplore.ieee.org/document/11353490}
}

@article{Scappucci2020,
  author  = {Scappucci, Giordano and Kloeffel, Christoph and Zwanenburg, Floris A. and Loss, Daniel and Myronov, Maksym and Zhang, Jian-Jun and De Franceschi, Silvano and Katsaros, Georgios and Veldhorst, Menno},
  title   = {The germanium quantum information route},
  journal = {Nature Reviews Materials},
  year    = {2020},
  volume  = {6},
  number  = {10},
  pages   = {926--943},
  doi     = {10.1038/s41578-020-00262-z}
}

@misc{Ademi2025,
  title         = {Distributing entanglement between distant semiconductor qubit registers using a shared-control shuttling link},
  author= {Ademi, Zarije and Bassi, Marion and Yu, C{\'e}cile X. and Oosterhout, Stefan D. and Matsumoto, Yuta and de Snoo, Sander L. and Sammak, Amir and Vandersypen, Lieven M. K. and Scappucci, Giordano and D{\'e}prez, Corentin and Veldhorst, Menno},
  year          = {2025},
  eprint        = {2510.26860},
  archivePrefix = {arXiv},
  primaryClass  = {cond-mat.mes-hall},
  note          = {arXiv:2510.26860}
}

@misc{Dijkema2026,
  title         = {Simultaneous operation of an 18-qubit modular array in germanium},
  author        = {Dijkema, J. J. and Zhang, X. and Bardakas, A. and Bouman, D. and
                   Cuzzocrea, A. and van Driel, D. and Girardi, D. and
                   Stehouwer, L. E. A. and Scappucci, G. and
                   Zwerver, A. M. J. and Hendrickx, N. W.},
  year          = {2026},
  eprint        = {2604.01063},
  archivePrefix = {arXiv},
  primaryClass  = {cond-mat.mes-hall},
  note          = {arXiv:2604.01063},
}

@article{Seidler2025,
  title         = {Spatial uniformity of g-tensor and spin-orbit interaction in germanium hole spin qubits},
  author        = {Seidler, Inga and Het{\'e}nyi, Bence and Sommer, Lisa and
                   Massai, Leonardo and Tsoukalas, Konstantinos and
                   Kelly, Eoin G. and Orekhov, Alexei and Aldeghi, Michele and
                   Bedell, Stephen W. and Paredes, Stephan and
                   Schupp, Felix J. and Mergenthaler, Matthias and
                   Salis, Gian and Fuhrer, Andreas and
                   Harvey-Collard, Patrick},
  year          = {2025},
  eprint        = {2510.03125},
  archivePrefix = {arXiv},
  primaryClass  = {cond-mat.mes-hall},
  note          = {arXiv:2510.03125}
}

@article{Nowack2007,
  author  = {Nowack, K. C. and Koppens, F. H. L. and Nazarov, Yu. V. and Vandersypen, L. M. K.},
  title   = {Coherent Control of a Single Electron Spin with Electric Fields},
  journal = {Science},
  volume  = {318},
  number  = {5855},
  pages   = {1430--1433},
  year    = {2007},
  doi     = {10.1126/science.1148092},
}

@article{Saraswat2006,
   author = {Krishna Saraswat and Chi On Chui and Tejas Krishnamohan and Donghyun Kim and Ammar Nayfeh and Abhijit Pethe},
   doi = {10.1016/j.mseb.2006.08.014},
   issn = {09215107},
   issue = {3},
   journal = {Materials Science and Engineering: B},
   month = {12},
   pages = {242-249},
   publisher = {Elsevier BV},
   title = {High performance germanium MOSFETs},
   volume = {135},
   year = {2006}
}

@article{Gupta2013,
   author = {Suyog Gupta and Robert Chen and James S. Harris and Krishna C. Saraswat},
   doi = {10.1063/1.4850518},
   issn = {00036951},
   issue = {24},
   journal = {Applied Physics Letters},
   month = {12},
   title = {Atomic layer deposition of Al2O3 on germanium-tin (GeSn) and impact of wet chemical surface pre-treatment},
   volume = {103},
   year = {2013}
}

@misc{PIN,
  author       = {Alexander Pawlis and Nils von den Driesch and Yurii Kutovyi and Benjamin Bennemann and Christoph Krause and Christine Falter},
  title        = {P-I-N - BAUTEIL UND VERFAHREN ZUR HERSTELLUNG},
  howpublished = {(WO2025168549)}
}

@misc{CONTACT,
  author       = {Alexander Pawlis and Janßen Johanna and Benjamin Bennemann and Christoph Krause},
  title        = {HERSTELLEN EINES OHMSCHEN KONTAKTS SOWIE ELEKTRONISCHES BAUELEMENT MIT OHMSCHEM KONTAKT},
  howpublished = {(WO2020212154)}
}

@book{Waser2003,
   author = {R. Waser},
   city = {Weinheim},
   isbn = {3527403639},
   publisher = {Wiley-VCH},
   title = {Nanoelectronics and Information Technology},
   year = {2003}
}

@article{Song2023,
   author = {Yiwen Song and Arkka Bhattacharyya and Anwarul Karim and Daniel Shoemaker and Hsien Lien Huang and Saurav Roy and Craig McGray and Jacob H. Leach and Jinwoo Hwang and Sriram Krishnamoorthy and Sukwon Choi},
   doi = {10.1021/acsami.2c21048},
   issn = {19448252},
   issue = {5},
   journal = {ACS Applied Materials and Interfaces},
   month = {2},
   pages = {7137-7147},
   pmid = {36700621},
   publisher = {American Chemical Society},
   title = {Ultra-Wide Band Gap Ga2O3-on-SiC MOSFETs},
   volume = {15},
   year = {2023}
}

@article{Duan2026,
   author = {Enchuan Duan and Yunfei Ma and Binju Qiu and Jinggui Zhou and Shuting Huang and Xia Xiao and Bo Zhang and Qi Zhou},
   doi = {10.1021/acsaelm.6c00225},
   issn = {26376113},
   issue = {12},
   journal = {ACS Applied Electronic Materials},
   month = {6},
   pages = {4953-4958},
   publisher = {American Chemical Society},
   title = {E-Mode GaN p-MOSFET Based on a Double Heterostructure with a Buried Back Gate Achieving Low Subthreshold Slope and Zero Hysteresis},
   volume = {8},
   year = {2026}
}

@article{Kang2023,
   author = {Junzhe Kang and Kai Xu and Hanwool Lee and Souvik Bhattacharya and Zijing Zhao and Zhiyu Wang and R. Mohan Sankaran and Wenjuan Zhu},
   doi = {10.1063/5.0134729},
   issn = {00036951},
   issue = {8},
   journal = {Applied Physics Letters},
   month = {2},
   publisher = {American Institute of Physics Inc.},
   title = {High Ion/Ioffratio 4H-SiC MISFETs with stable operation at 500 °C using SiO2/SiNx/Al2O3gate stacks},
   volume = {122},
   year = {2023}
}

@article{Kaul2025,
   author = {Prateek Kaul and Jan Karthein and Jonas Buchhorn and Taizo Kawano and Taisei Usubuchi and Jun Ishihara and Nicolas Rotaru and Patrick Del Vecchio and Omar Concepcion and Zoran Ikonic and Detlev Grützmacher and Qing Tai Zhao and Oussama Moutanabbir and Makoto Kohda and Thomas Schäpers and Dan Buca},
   doi = {10.1038/s43246-025-00934-9},
   issn = {26624443},
   issue = {1},
   journal = {Communications Materials},
   month = {12},
   publisher = {Springer Nature},
   title = {GeSn quantum wells as a platform for spin-resolved hole transport},
   volume = {6},
   year = {2025}
}

@article{Lodari2021,
   author = {Mario Lodari and Nico W. Hendrickx and William I.L. Lawrie and Tzu Kan Hsiao and Lieven M.K. Vandersypen and Amir Sammak and Menno Veldhorst and Giordano Scappucci},
   doi = {10.1088/2633-4356/abcd82},
   issn = {26334356},
   issue = {1},
   journal = {Materials for Quantum Technology},
   month = {3},
   publisher = {Institute of Physics},
   title = {Low percolation density and charge noise with holes in germanium},
   volume = {1},
   year = {2021}
}

@article{Lyu1993,
   author = {Jong-Son Lyu},
   doi = {10.4218/etrij.93.0193.0002},
   issn = {1225-6463},
   issue = {2},
   journal = {ETRI Journal},
   month = {10},
   pages = {10-25},
   publisher = {Wiley-Blackwell},
   title = {A New Method for Extracting Interface Trap Density in Short-Channel MOSFETs from Substrate-Bias-Dependent Subthreshold Slopes},
   volume = {15},
   year = {1993}
}

@article{Elzerman2004,
   author = {J. M. Elzerman and R. Hanson and L. H. Willems van Beveren and B. Witkamp and L. M. K. Vandersypen and L. P. Kouwenhoven},
   doi = {10.1038/nature02693},
   issn = {00280836},
   issue = {6998},
   journal = {Nature},
   month = {7},
   pages = {429-431},
   title = {An X-ray outburst from the rapidly accreting young star that illuminates McNeil's nebula},
   volume = {430},
   year = {2004}
}

@article{Massai2024,
   author = {Leonardo Massai and Bence Hetényi and Matthias Mergenthaler and Felix J. Schupp and Lisa Sommer and Stephan Paredes and Stephen W. Bedell and Patrick Harvey-Collard and Gian Salis and Andreas Fuhrer and Nico W. Hendrickx},
   doi = {10.1038/s43246-024-00563-8},
   issn = {26624443},
   issue = {1},
   journal = {Communications Materials},
   month = {12},
   publisher = {Springer Nature},
   title = {Impact of interface traps on charge noise and low-density transport properties in Ge/SiGe heterostructures},
   volume = {5},
   year = {2024}
}

@article{Park2021,
   author = {Youngseo Park and Jiyeon Ma and Geonwook Yoo and Junseok Heo},
   doi = {10.3390/nano11020494},
   issn = {20794991},
   issue = {2},
   journal = {Nanomaterials},
   month = {2},
   pages = {1-10},
   publisher = {MDPI AG},
   title = {Interface trap-induced temperature dependent hysteresis and mobility in $\beta$-Ga2O3 field-effect transistors},
   volume = {11},
   year = {2021}
}
